\documentclass[twocolumn,tighten,twocolappendix]{aastex63}

\usepackage{amssymb,amsmath,amsthm}
\usepackage{graphicx}
\usepackage{mathrsfs}
\usepackage{booktabs}
\usepackage{multirow}
\usepackage{empheq}
\usepackage{verbatim}
\usepackage{mathtools}
\usepackage{hyperref}
\usepackage{xcolor}
\usepackage{xspace}
\usepackage{lineno}
\usepackage[makeroom]{cancel}
\usepackage[acronym]{glossaries}
\usepackage{CJKutf8}

\def\j0958{J0958\ensuremath{+}6533\xspace} % J0958+6533
\def\oj287{OJ\,287\xspace} % OJ 287

\newcommand{\chinese}[1]{\begin{CJK*}{UTF8}{gbsn}(#1)\end{CJK*}}

\makeatletter
\DeclareRobustCommand{\pder}[1]{%
  \@ifnextchar\bgroup{\@pder{#1}}{\@pder{}{#1}}}
\newcommand{\@pder}[2]{\frac{\partial#1}{\partial#2}}
\makeatother

\makeatletter
\DeclareRobustCommand{\ppder}[1]{%
  \@ifnextchar\bgroup{\@ppder{#1}}{\@ppder{}{#1}}}
\newcommand{\@ppder}[2]{\frac{\partial^2#1}{\partial#2^2}}
\makeatother

\newcommand{\haystack}{Massachusetts Institute of Technology, Haystack Observatory, 99 Millstone Rd, Westford, MA 01886, USA}
\newcommand{\bhi}{Black Hole Initiative, Harvard University, 20 Garden Street, Cambridge, MA 02138, USA}
\newcommand{\asiaa}{Institute of Astronomy and Astrophysics, Academia Sinica, 11F of Astronomy-Mathematics Building, AS/NTU No. 1, Sec. 4, Roosevelt Road, Taipei 106216, Taiwan, R.O.C.}
\newcommand{\cfa}{Center for Astrophysics $|$ Harvard \& Smithsonian, 60 Garden Street, Cambridge, MA 02138, USA}
\newcommand{\mpifr}{Max-Planck-Institut f{\"u}r Radioastronomie, Auf dem H{\"u}gel 69, D-53121 Bonn, Germany}
\newcommand{\arizona}{University of Arizona, 933 North Cherry Avenue, Tucson, AZ 85721, USA}
\newcommand{\kasi}{Korea Astronomy and Space Science Institute, Daedeok-daero 776, Yuseong-gu, Daejeon 34055, Korea}
\newcommand{\iram}{Institut de Radioastronomie Millim\'etrique (IRAM), Avenida Divina Pastora 7, Local 20, E-18012, Granada, Spain}
\newcommand{\france}{Institut de Radioastronomie Millim\'etrique (IRAM), 300 rue de la Piscine, F-38406 Saint Martin d’H\`eres, France}
\newcommand{\shanghai}{Shanghai Astronomical Observatory, Chinese Academy of Sciences, 80 Nandan Road, Shanghai 200030, People's Republic of China}
\newcommand{\icrar}{International Centre for Radio Astronomy Research, M468, The University of Western Australia, 35 Stirling Hwy, Perth, WA 6009, Australia}

\newcommand{\ign}{Observatorio Astron{\'o}mico Nacional (IGN), Alfonso XII, 3 y 5, 28014 Madrid, Spain}

\newcommand{\malaya}{Department of Physics, Faculty of Science, Universiti Malaya, 50603 Kuala Lumpur, Malaysia}

\begin{document}

\title{First frequency phase transfer from the 3\,mm to the 1\,mm band on an Earth-sized baseline}

\correspondingauthor{Sara~Issaoun}
\email{sara.issaoun@cfa.harvard.edu}

\author[0000-0002-5297-921X]{Sara Issaoun}
\affil{\cfa}\affil{\bhi}
\author[0000-0002-5278-9221]{Dominic W. Pesce}
\affil{\cfa}\affil{\bhi}
\author[0000-0003-4871-9535]{Mar\'ia J. Rioja}
\affil{\icrar}\affil{\ign}
\author[0000-0003-0392-3604]{Richard Dodson}
\affil{\icrar}
\author[0000-0002-9030-642X]{Lindy Blackburn}
\affil{\cfa}\affil{\bhi}
\author[0000-0002-3490-146X]{Garrett K. Keating}
\affil{\cfa}
\author[0000-0002-9031-0904]{Sheperd S. Doeleman}
\affil{\cfa}\affil{\bhi}
\author[0000-0002-4148-8378]{Bong Won Sohn}
\affil{\kasi}
\author[0000-0001-7369-3539]{Wu Jiang \chinese{江悟}}
\affil{\shanghai}
\author{Dan Hoak}
\affil{\haystack}
\author[0000-0002-5168-6052]{Wei Yu \chinese{于威}}
\affil{\cfa}
\author[0000-0001-8700-6058]{Pablo Torne}
\affil{\iram}
\author[0000-0002-1407-7944]{Ramprasad Rao}
\affil{\cfa}
\author[0000-0002-6514-553X]{Remo P. J. Tilanus}
\affil{\arizona}
\author[0000-0003-3708-9611]{Iv\'an Mart\'i-Vidal}
\affil{Departament d'Astronomia i Astrof\'isica, Universitat de Val\`encia, C. Dr. Moliner 50, E-46100 Burjassot, Val\`encia, Spain}
\affil{Observatori Astron\`omic, Universitat de Val\`encia, C. Catedrático Jos\'e Beltr\'an 2, E-46980 Paterna, Val\`encia, Spain}
\author[0000-0001-7003-8643]{Taehyun Jung}
\affil{\kasi}
\author{Garret Fitzpatrick}
\affil{\cfa}
\author[0000-0003-0981-9664]{Miguel S\'anchez-Portal}
\affil{\iram}
\author[0000-0002-8042-5951]{Salvador S\'anchez}
\affil{\iram}
\author[0000-0002-4603-5204]{Jonathan Weintroub}
\affil{\cfa}
\author[0000-0003-0685-3621]{Mark Gurwell}
\affil{\cfa}
\author[0000-0002-4908-4925]{Carsten Kramer}
\affil{\france}
\author[0000-0001-7622-3890]{Carlos Dur\'an}
\affil{\iram}
\author{David John}
\affil{\iram}
\author{Juan L. Santaren}
\affil{\iram}
\author{Derek Kubo}
\affiliation{Institute of Astronomy and Astrophysics, Academia Sinica, 645 N. A'ohoku Place, Hilo, HI 96720, USA}
\author{Chih-Chiang Han}
\affil{\asiaa}
\author[0000-0003-1799-8228]{Helge Rottmann}
\affil{\mpifr}
\author[0000-0003-1938-0720]{Jason SooHoo}
\affil{\haystack}
\author[0000-0002-7128-9345]{Vincent L. Fish}
\affil{\haystack}
\author[0000-0002-4417-1659]{Guang-Yao Zhao}
\affil{\mpifr}
\author[0000-0001-6993-1696]{Juan Carlos Algaba}
\affil{\malaya}
\author[0000-0002-7692-7967]{Ru-Sen Lu \chinese{路如森}}
\affil{\shanghai}
\affil{Key Laboratory of Radio Astronomy and Technology, Chinese Academy of Sciences, A20 Datun Road, Chaoyang District, Beijing, 100101, People’s Republic of China}
\affil{\mpifr}
\author[0000-0001-6083-7521]{Ilje Cho}
\affil{Department of Astronomy, Yonsei University, Yonsei-ro 50, Seodaemun-gu, 03722 Seoul, Republic of Korea}
\affil{\kasi}
\author[0000-0002-2127-7880]{Satoki Matsushita}
\affil{\asiaa}
\author[0000-0003-2890-9454]{Karl-Friedrich Schuster}
\affil{\france}

\begin{abstract}
Frequency Phase Transfer (FPT) is a technique designed to increase coherence and sensitivity in radio interferometry by making use of the non-dispersive nature of the troposphere to calibrate high-frequency data using solutions derived at a lower frequency. While the Korean VLBI Network has pioneered the use of simultaneous multi-band systems for routine FPT up to an observing frequency of 130\,GHz, this technique remains largely untested in the (sub)millimeter regime.  A recent effort has been made to outfit dual-band systems at (sub)millimeter observatories participating in the Event Horizon Telescope (EHT) and to test the feasibility and performance of FPT up to the observing frequencies of the EHT. We present the results of simultaneous dual-frequency observations conducted in January 2024 on an Earth-sized baseline between the IRAM 30-m in Spain and the JCMT and SMA in Hawai`i. We performed simultaneous observations at 86 and 215\,GHz on the bright sources J0958+6533 and OJ\,287, with strong detections obtained at both frequencies. We observe a strong correlation between the interferometric phases at the two frequencies, matching the trend expected for atmospheric fluctuations and demonstrating for the first time the viability of FPT for VLBI at a wavelength of $\sim$1\,millimeter. We show that the application of FPT systematically increases the 215\,GHz coherence on all averaging timescales. In addition, the use of the co-located JCMT and SMA as a single dual-frequency station demonstrates the feasibility of paired-antenna FPT for VLBI for the first time, with implications for future array capabilities (e.g., ALMA sub-arraying and ngVLA calibration strategies).
\end{abstract}

\keywords{galaxies: active -- galaxies: jets -- techniques: interferometric -- very long baseline interferometry}

% \tableofcontents

\section{Introduction}

Frequency Phase Transfer (FPT) is a technique to increase coherence and sensitivity in interferometric observations via the use of phase, delay, and rate solutions from lower frequencies to calibrate higher frequencies \citep[see][and references therein]{Rioja_2020}. Optimal FPT requires simultaneous observations of a source at two or more different frequencies along the same line-of-sight or optical path, and it requires that the source be detected at the lower frequency within the coherence timescale at the highest frequency, such that phase variations within this timescale can be tracked at the lower frequency. Provided that the observing equipment is well-calibrated and stable, these phase variations originate in the troposphere and are primarily non-dispersive, meaning that the magnitude of the variations is proportional to the observing frequency. A scaled-up version of the lower frequency phase solution can thus be used to calibrate the higher frequency data, with the potential to substantially increase the coherent integration time -- and therefore sensitivity -- at the higher frequency.

The potential benefit of FPT for calibrating high-frequency radio observations has been recognized for several decades, and multiple different strategies have been devised and employed within the context of connected-element interferometers.  The most direct application of FPT requires simultaneous observations at multiple frequencies with each antenna in the array, as demonstrated by, e.g., the Submillimeter Array \citep[SMA;][]{Masson_1989,Hunter_2005}.  But FPT-enabled calibration improvements can also be realized using a ``paired-antenna'' mode in which nearby telescopes observe simultaneously at two different frequencies, such as the ``paired antennas method'' used by the Nobeyama Millimeter Array \citep{Asaki_1996,Asaki_1998} and the ``paired antenna calibration system'' used by the Combined Array for Research in Millimeter-wave Astronomy \citep[CARMA;][]{Perez_2010,Zauderer_2016}.  Another alternative is to conduct fast frequency-switching observations, such as the ``band-to-band'' calibration used at the Atacama Large Millimeter/submillimeter Array \citep[ALMA;][]{Asaki_2020a,Asaki_2020b,Maud_2020,Maud_2022,Maud_2023,Asaki_2023}.

The FPT approach was first used in very long baseline interferometry (VLBI) by \citet{Middelberg_2005} to increase the coherence time at 86\,GHz for the Very Long Baseline Array (VLBA). Work by M.~Rioja and R.~Dodson \citep{Dodson_2009,Rioja_2011} pioneered FPT techniques using the VLBA up to 86\,GHz including frequency and source switching, enabling precise astrometry. Subsequently, FPT applied to simultaneous Korean VLBI Network \citep[KVN,][]{Han_2013,Rioja_2015} observations at 22, 43, 87, and 130\,GHz increased coherence times at 130\,GHz from tens of seconds to $\sim$20 minutes, and dual-band intercontinental VLBI at 22 and 43\,GHz between KVN Ulsan and the Yebes 40m telescope in Spain was demonstrated in \cite{Jung_2015}. FPT has been successful for other KVN programs, such as the Interferometric MOnitoring of GAmma-ray Bright AGN (iMOGABA) where it has enabled the imaging of several sources at 86 and 129\,GHz that were not detected without FPT \citep{Algaba2015}. In \citet{Zhao2019}, FPT was applied to simultaneous 22 and 43\,GHz observations with the KVN and VERA in Japan (combined as KaVA), which increased the coherence time at 43\,GHz from $\sim$1 minute to tens of minutes. Such applications at lower frequencies have demonstrated that coherence times can be extended to tens of minutes with the help of FPT \citep{Rioja_2014}, and extended to multiple hours with the addition of a third frequency \citep{Zhao_2018} and/or another source \citep{Rioja_2011} to remove residual ionospheric and instrumental terms. The Event Horizon Telescope (EHT) has also achieved integration times of minutes at 230\,GHz using phase stabilization for sources with flux densities ${\gtrsim}$1\,Jy \citep{M87PaperIII}. Prospects of FPT with an added lower frequency band would enable similarly increased coherence times for substantially weaker sources than would otherwise be observable with the EHT, and would significantly improve the ability to observe at 345\,GHz more routinely \citep{Issaoun_2023, Rioja_2023,Pesce_2024,Raymond_2024}.

The FPT technique is well exercised at frequencies up to 130\,GHz but has remained untested at higher frequencies on VLBI baselines, largely due to the lack of (sub)millimeter facilities with VLBI-capable dual-band setups.  Motivation for developing the FPT capability has been building in recent years, as higher-frequency VLBI observations form a cornerstone of near-term \citep{EHT_midrange} and next-generation EHT science goals \citep{ngEHT_reference_array}.
The Institut de Radioastronomie Millim\'etrique 30\,m telescope (IRAM 30-m) is one of the first observatories
capable of observing simultaneously at two frequencies in VLBI mode in the (sub)millimeter regime. This capability is the result of recent in-house modifications to the existing Eight Mixer Receiver \citep[EMIR;][]{Carter_2012} currently in operation at the IRAM 30-m.

In this paper, we describe the first successful demonstration of FPT between 86 and 215\,GHz on an Earth-sized VLBI baseline. We describe the observations in \autoref{sec:obs}, correlation and calibration of the data in \autoref{sec:data}, and the comparison of single-frequency and FPT analysis in \autoref{sec:analysis}. A discussion of the results is presented in \autoref{sec:discussion}, and we conclude and summarize in \autoref{sec:Summary}.

\begin{figure}
    \centering
    \includegraphics[width=1.0\linewidth]{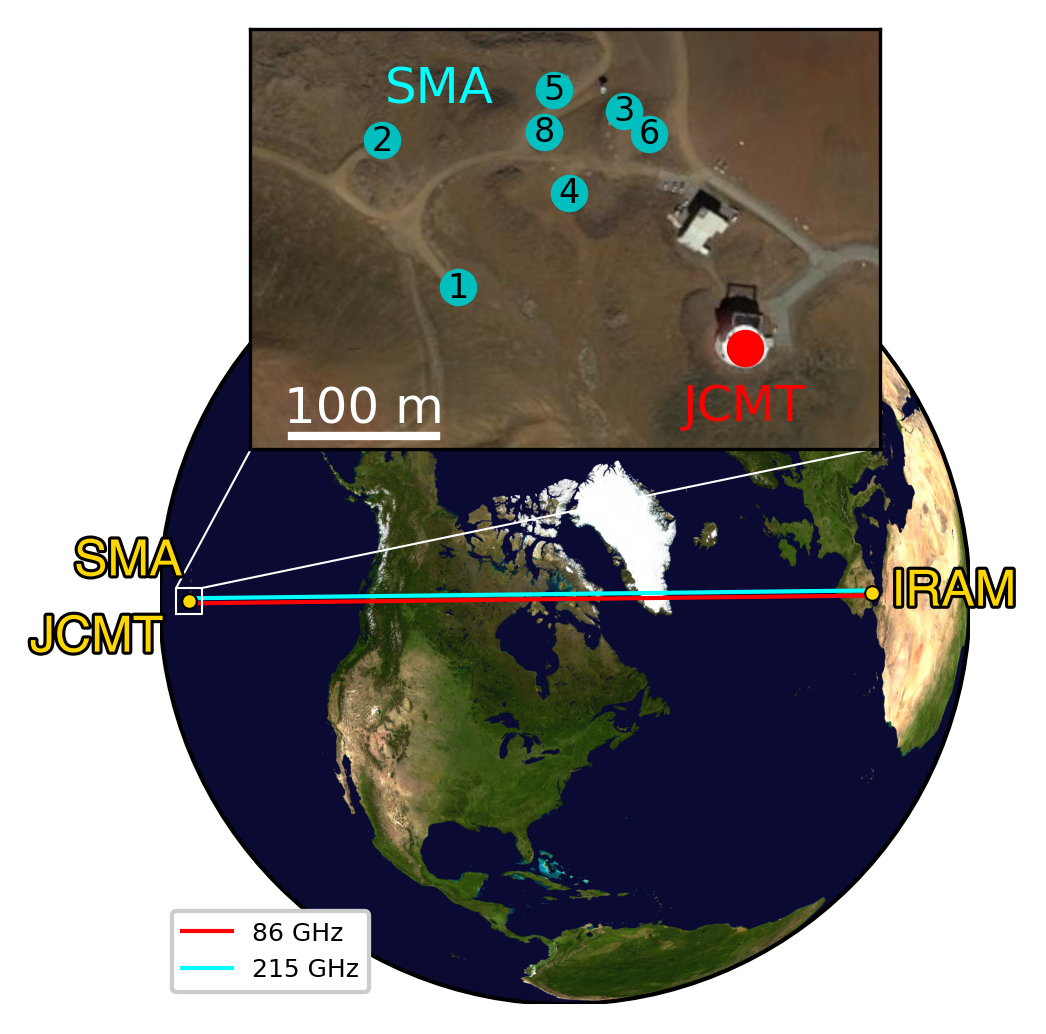}
    \caption{The array used for the observations presented in this paper, as seen from the source \j0958. The IRAM$-$JCMT baseline at 86\,GHz is shown in red, the IRAM$-$SMA baseline at 215\,GHz is shown in cyan. The inset on top shows the distribution of antennas on Maunakea, with the JCMT shown in red and the individual SMA antennas shown in cyan and enumerated. Antenna 6 was the SMA reference antenna for our observations.}
    \label{fig:globe}
\end{figure}

\section{Observations}\label{sec:obs}

We carried out simultaneous 86 and 215\,GHz VLBI observations on 2024 January 24. The participating observatories were the James Clerk Maxwell Telescope (JCMT) and the phased SMA on Maunakea in Hawai`i, and the IRAM 30-m telescope (IRAM) on Pico Veleta in Spain (see \autoref{fig:globe}).  The sources \oj287, \j0958, and J1033$+$6051 were interleaved across nine scans.
Each scan was seven minutes in duration, with a total observing track length of $\sim$2 hours. \autoref{tab:Schedule} describes the observing schedule. Observations took place following a go/no-go decision based on weather conditions and technical readiness during the potential trigger window of 2024 January 23-25. Weather was good at IRAM (optical depth $\tau_{225} \sim 0.25$) and excellent on Maunakea (optical depth $\tau_{225} \sim 0.05$).

The VLBI data were recorded in two polarizations onto Mark6 recorders \citep{whitney} via ROACH2 digital backends \citep{Vertatschitsch_2015} at the JCMT and SMA and a Digital Broadband Backend 3 \citep[DBBC3;][]{Tuccari_2014} at the IRAM 30-m. The observatories recorded in 2\,GHz-wide frequency bands: 86-88\,GHz for the lower frequency and 214-216\,GHz for the higher frequency. The JCMT and SMA were observing as a paired dual-frequency antenna in circular polarization basis, with the JCMT at 86\,GHz and the SMA at 215\,GHz. The physical spatial separation between the two telescopes is $\sim$160\,m, see \autoref{fig:globe} for the antenna distribution on Maunakea. IRAM observed at both 86 and 215\,GHz simultaneously. Figure~\ref{fig:globe} illustrates the baselines in this experiment at both frequencies. Due to the optics setup at IRAM, quarter-waveplates could not be inserted for both frequency bands simultaneously, thus its native linear polarization basis was maintained.

Because the scaling of phase variations at the lower frequency can result in phase-wrapping ambiguities at the higher frequency, it is preferable to select frequencies such that the frequency ratio between the lower and higher frequencies is an integer \citep{Rioja_2020}. For the more general non-integer case, these ambiguities can introduce seemingly random phase jumps whenever the lower frequency phase wraps, and a clean unwrapping of the phases becomes increasingly difficult as the signal-to-noise ratio (S/N) decreases \citep[see discussions in][]{Dodson_2014}. However, the frequency ratio for this experiment is 2.5, a non-integer. This setup was selected in order to replicate frequency tunings regularly used by the observatories for EHT and Global Millimeter VLBI Array observing campaigns, which both reduces risks associated with ad hoc scheduling of individual tests and simplified the switch to standard EHT setup for the EHT dress rehearsal tests that were taking place concurrently to this FPT test. 

Following the observing run, the data were transferred to correlator facilities for correlation. We selected three high-priority scans for data transfer and analysis: two scans on \j0958 and one scan on \oj287. The SMA and IRAM priority scans were e-transferred shortly following the observations, while the JCMT data were shipped later with the recording media from the 2024 April EHT science campaign.  One of the disks of a JCMT module was found to be corrupted, however this did not impact the priority scans. An estimated $\sim$25\% data loss was identified in the IRAM 86\,GHz data for the vertical linear polarization data, which spreads uniformly across scans and is likely attributed to an issue in the recording equipment. Subsequent disk pack conditioning at the SMA erased the untransferred data, so only the three initially-transferred scans were available for correlation; see \autoref{tab:Schedule} for details.

\begin{deluxetable*}{lcccccccc}
\tablecolumns{9}
\tablewidth{0pt}
\tablecaption{Observing schedule for 2024 January 24 \label{tab:Schedule}}
\tablehead{\colhead{Scan number} & \colhead{Target} & \colhead{Start time } & \colhead{Duration } & \colhead{Hawai`i elevation } & \colhead{IRAM elevation } & \colhead{Retained?} & \colhead{S/N } & \colhead{S/N }\\
& & \colhead{(UT)} & \colhead{(s)} & \colhead{(deg)} & \colhead{(deg)} &  & \colhead{(86\,GHz)} & \colhead{(215\,GHz)}}
\startdata
01 & \oj287       & 05:45:00 & 420 & 16.2 & 25.4 & no & - & - \\
02 & \oj287       & 06:00:00 & 420 & 19.6 & 22.4 & no & - & - \\
03 & \oj287       & 06:15:00 & 420 & 23.1 & 19.5 & yes & 928 & 141 \\
04 & \j0958       & 06:30:00 & 420 & 20.4 & 42.3 & no & - & - \\
05 & \j0958       & 06:45:00 & 420 & 21.9 & 40.7 & yes & 3579 & 992 \\
06 & \j0958       & 07:00:00 & 420 & 23.5 & 39.2 & no & - & - \\
07 & J1033$+$6051 & 07:09:00 & 420 & 20.6 & 41.7 & no & - & - \\
08 & \j0958       & 07:18:00 & 420 & 25.4 & 37.3 & yes & 3484 & 904 \\
09 & J1033$+$6051 & 07:33:00 & 420 & 23.5 & 38.8 & no & - & - \\
\enddata
\tablecomments{Properties of each scan in the observing track carried out on 2024 January 24.  From left to right, the columns list the scan number, the target source, the scan start time in UT, the duration of the scan in seconds, and whether the scan was retained after e-transferring the data. S/N for processed scans refer to the S/N of the full-band (2\,GHz), scan-average correlation amplitude (incoherently averaged using 2-second segmentation) and summed in quadrature over all four polarization products.}
\end{deluxetable*}

\section{Data processing}\label{sec:data}

\subsection{Correlation}

Because the SMA SWARM correlator architecture outputs SMA data in the frequency domain in a larger frequency range compared to other sites \citep{Young_2016}, an offline pre-processing pipeline is used to filter, frequency-convert, and output data in the time domain. This pipeline, called the Adaptive Phased-array and
Heterogeneous Interpolating Downsampler for SWARM \citep[APHIDS,][]{Primiani_2016}, ensures that the SMA data delivered to the VLBI correlator matches that of the single-dish stations. The APHIDS pre-processing was carried out at the Smithsonian Astrophysical Observatory for the priority scans before e-transferring the SMA data to the correlator facilities. 

The data were correlated at both the MIT Haystack Observatory and the Shanghai Astronomical Observatory in parallel, with both computer clusters running the \textit{DiFX} software package \citep{Deller_2011}. Data were correlated with an accumulation period of 0.4\,s and a frequency resolution of 0.5\,MHz. The frequency ranges between the Hawai`i stations and IRAM do not have an exact overlap. \textit{DiFX outputbands} \citep{jan_wagner_2020_4319257} is used to synthesize 32 contiguous digital IF sub-bands spacing 62\,MHz and 62.5\,MHz for 215\,GHz and 86\,GHz respectively. Due to the different polarization setups at the participating sites, the output from the correlator is in a mixed-polarization basis. While a conversion to a matching polarization basis could be done to recover S/N lost in mixed basis \citep[with e.g., \textit{PolConvert};][]{martividal2016}, it was not deemed critical for the analysis as we are not S/N limited for this demonstration. At the correlator, fringes on \j0958 at 215\,GHz between IRAM and SMA were recovered at S/N$\sim$200, and fringes at 86\,GHz between IRAM and JCMT were recovered at S/N$\sim$800. No major issues were found at correlation, and both correlators produced identical outputs following some minor fixes from incorrect logging of station setups. To avoid duplication of efforts, only the correlator output from Haystack was used for subsequent analysis. 
The correlator output from \textit{DiFX} was converted from the Swinburne format to \textit{Mark4} format for processing through the Haystack Observatory Postprocessing System \citep[\textit{HOPS},][]{Whitney_2004,lindyhops}, and to \textit{FITS-IDI} for further processing with the NRAO Astronomical Image Processing System \citep[\textit{AIPS},][]{aips,greisen2011fits}.

\subsection{Calibration} \label{sec:calibration}

Three seven-minute scans of data are available for analysis (see \autoref{tab:Schedule}), but Scan 08 on \j0958 suffers from apparent instrumental corruptions that preclude a demonstration of the FPT technique.  With only a single baseline, we have limited options for mitigating or even uniquely identifying such data issues, so for this paper we present only the remaining two scans (Scan 03 on \oj287 and Scan 05 on \j0958).

We carry out the VLBI data analysis using multiple independent methods for single-frequency and dual-frequency FPT calibration, including approaches that use both standard softwares as well as more custom routines.  The dataset is relatively simple -- consisting of two scans on bright targets with only a single geometric baseline -- and we generally find excellent quantitative agreement between the output of all calibration methods.  To avoid excessive redundancy, we thus report calibration details for only two of the methods: \textit{HOPS} and \textit{AIPS}.  Unless otherwise specified, all results presented here correspond to data that were calibrated using the \textit{HOPS} method.

\subsubsection{\texttt{HOPS}}

The \textit{HOPS} \textit{fourfit} routine is used to fit delay and delay-rate solutions to the correlator \textit{Mark4} output and to further average down the data.  A ``single-band delay'' is used to average visibilities within each IF sub-band, followed by a ``multi-band delay'' parameter for averaging over the entire 2\,GHz band.  Independent delay and delay-rate parameters are fit for each baseline and polarization product, which accommodates an uncalibrated relative delay across polarization feeds.  Relative delays and delay-rates between 215\,GHz and 86\,GHz are also left unconstrained, which allows for uncalibrated residual instrumental delays and residual relative delay-rates that could arise from, e.g., small errors in antenna position.

An amplitude bandpass correction is applied (at the \textit{Mark4} conversion stage) by normalizing by the autocorrelation amplitude in each sub-band. Following an initial pass through the data, the strong baseline detections are also used to derive a phase bandpass correction that is piecewise linear across each sub-band and constant throughput the experiment. After the bandpass phase has been flattened, refined fringe solutions for delay and delay-rate are obtained and the corrected visibilities are averaged over the entire band.  These fringe solutions are referenced to the band center, which is 215\,GHz for the high-frequency band and 87\,GHz for the low-frequency band.  The fitted delay-rate model is also removed from the data, but data are kept at their original 0.4\,s resolution so that residual non-linear time-dependence in the phase structure can be further analyzed.

The processing steps described above are accomplished through the application of the \textit{EHT-HOPS} pipeline \citep{lindyhops} up to Stage~2 (phase bandpass); downstream stages of the pipeline that normally solve for stochastic phase variability, relative feed delays, and fringe closure are not used for this single-baseline experiment.

\begin{figure*}[t]
    \centering
    \includegraphics[width=1.00\linewidth]{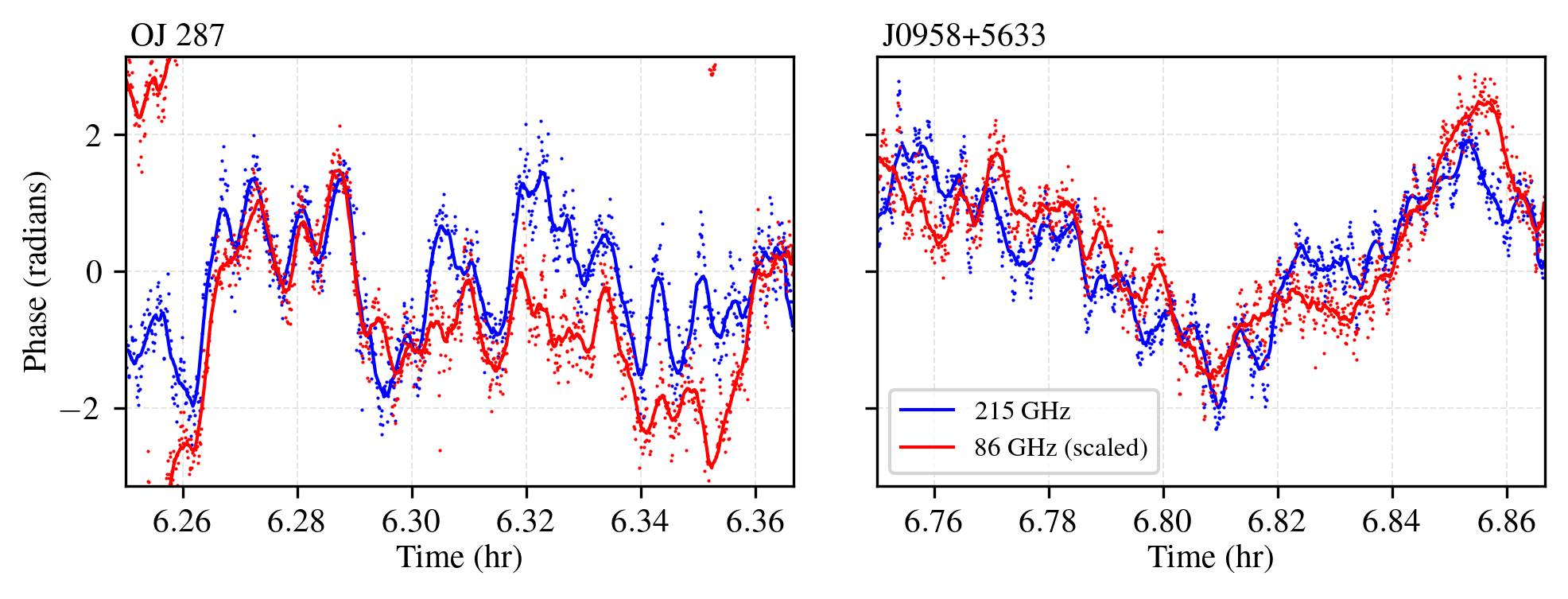}
    \caption{Bandpass-calibrated and rate-removed visibility phases on the IRAM$-$JCMT baseline at 86\,GHz after scaling by the frequency ratio (in red) and on the IRAM$-$SMA baseline at 215\,GHz (in blue), with moving 10\,s averages overlaid as solid curves. The phases are shown for Scan 03 on the source \oj287 (left) and for Scan 05 on the source \j0958 (right). Time is indicated in UT hours on 2024 January 24.
    }
    \label{fig:best-scan}
\end{figure*}

\subsubsection{\texttt{AIPS}}

The \textit{AIPS} \citep{aips} calibration approach largely follows standard procedures for calibration of single-frequency VLBI datasets, followed by the somewhat less standard procedures for dual-frequency FPT analysis \citep{Rioja_2011,Dodson_2014}, which we outline below. We first ensured that the individual IF sub-bands at both frequencies are aligned across the 2\,GHz bandwidth and thus can be frequency-averaged by removing the instrumental terms. We use the \textit{AIPS} task \textit{FRING} to solve for the independent phase and delay values for each IF and for each polarization over a time interval of 30 seconds.  These values are applied to all scans of the corresponding dataset, a procedure known as the ``manual phase calibration" step.

To measure the coherence in the visibility time series 
we use \textit{FRING} to solve for phase, delay, and rate over the whole bandwidth and both polarizations, using ten different solution intervals ranging from 1 to 500 seconds, over the entire duration of Scans 03 and 05. 
This step is done independently at 86 and 215\,GHz for the single-frequency analysis, which outputs time-series of visibilities at the original 0.4\,s resolution.

For the dual-frequency FPT analysis we use the 86\,GHz dataset to precondition the 215\,GHz dataset prior to the coherence measurement. After the manual phase calibration, we use \textit{FRING} to solve for phase, delay, and rate on the 86\,GHz frequency-averaged dataset over the entire duration of the two scans with a short solution interval (10\,s on \oj287 and 2.4\,s on \j0958).  
These solutions are transferred and applied to the 215\,GHz dataset, after properly scaling by the frequency ratio, to provide the pre-conditioning for the dual-frequency FPT analysis.

\section{Analysis} \label{sec:analysis}

The output of the calibration procedures described in \autoref{sec:calibration} is a time series of visibility phases at 86 and 215\,GHz that have each been coherently averaged in frequency across their respective bands.  \autoref{fig:best-scan} shows these single-frequency time series for both Scan 03 on \oj287 and Scan 05 on \j0958.  A qualitative similarity between the two phase time series is visually apparent for both scans.  In this section, we conduct analyses that aim to assess the degree of similarity more quantitatively.  To maximize S/N, we average over all correlation products prior to carrying out the analyses.

\subsection{Correlated phases}

\begin{figure}[t]
    \centering
    \includegraphics[width=1.00\linewidth]{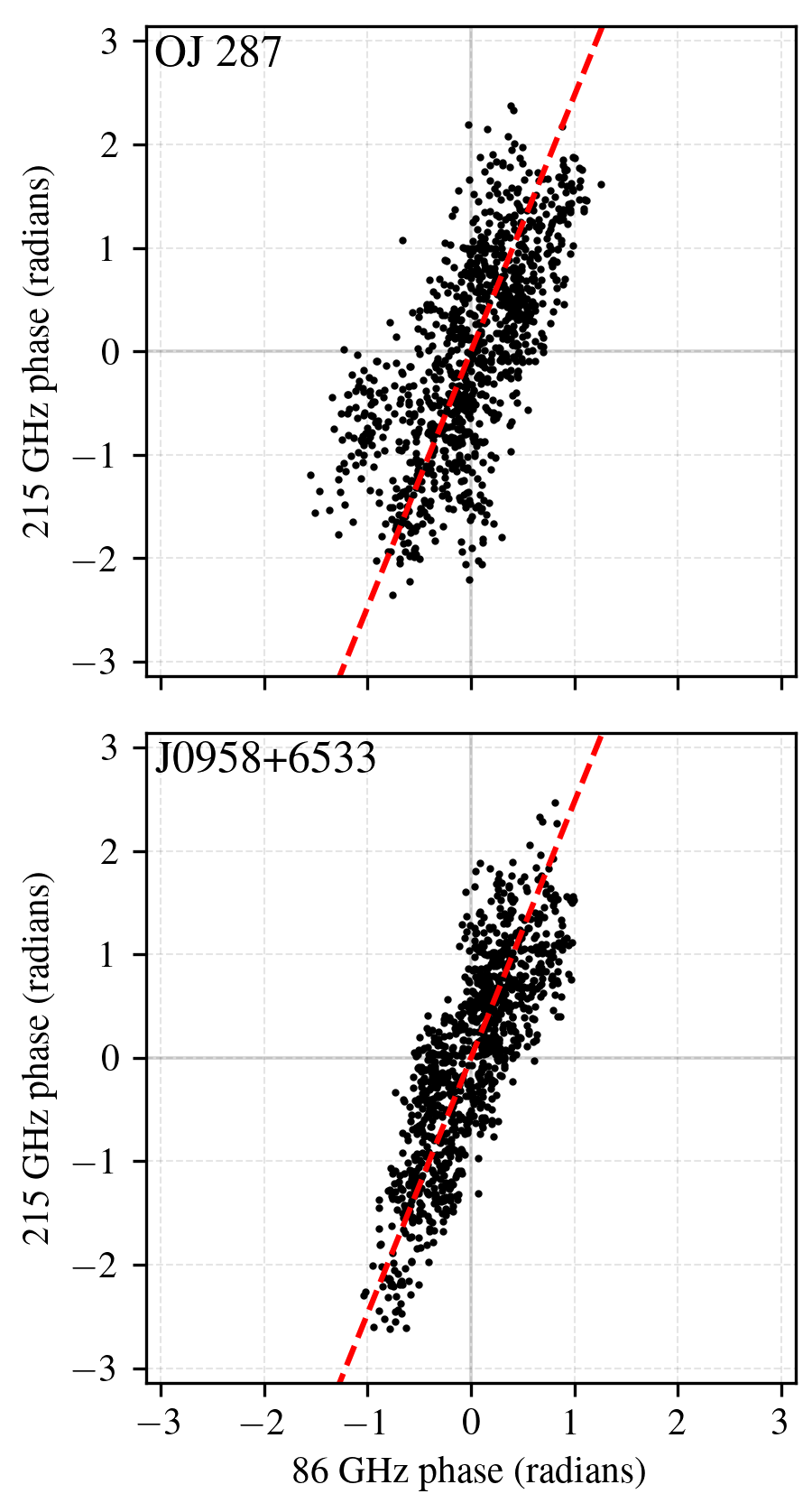}
    \caption{Correlation between 86 and 215\,GHz visibility phases for \oj287 (top) and \j0958 (bottom).  The dashed red line in both panels has a slope equal to the frequency ratio of $\sim$2.5.}
    \label{fig:correlation}
\end{figure}

\autoref{fig:correlation} shows the correlation between the 86 and 215\,GHz visibility phases for both targets.  We find that, in both cases, the phases at the two observing bands are correlated along a direction consistent with their frequency ratio, as expected for non-dispersive fluctuations. The Pearson correlation coefficient between the 86 and 215\,GHz phases for \oj287 is $\sim$0.68, while for \j0958 it is $\sim$0.81.

\subsection{Coherence}

Practically, we are most interested in FPT because of its potential to extend coherent integration times for high-frequency observations.  We can use our observed phase time-series to assess the quality of coherence improvement that FPT with this dataset would achieve.  For a phase time-series $\phi(t)$, we compute the coherence $C$ as

\begin{equation}
C(T) = \frac{e^{\sigma^2/2}}{N-T+1} \sum_{j=1}^{N-T+1} \left| \frac{1}{T} \sum_{k=j}^{j+T-1} e^{i \phi(t_k)} \right| , \label{eqn:Coherence}
\end{equation}

\noindent where the inner sum describes a coherent average of the phasor $e^{i\phi}$ over $T$ consecutive data points (starting at data point $j$), the outer sum describes an incoherent average over all $N-T+1$ segments of data that contain $T$ consecutive data points, $N$ is the total number of data points in the dataset, and $\sigma$ is the thermal noise level in the phases.  The prefactor $e^{\sigma^2/2}$ serves to normalize the coherence to unity in the presence of only thermal noise \citep[e.g.,][Eq.\,13.94]{TMS}.
We estimate the thermal noise level by computing the standard deviation of the differences of consecutive pairs of data points,

\begin{equation}
\sigma = \frac{1}{\sqrt{2}} \sqrt{\left\langle \big[ \phi(t_{k+1}) - \phi(t_k) \big]^2 \right\rangle_k} \;, \label{eqn:ThermalNoise} 
\end{equation}

\noindent where the notation $\langle \cdot \rangle_k$ indicates an average taken over all sampled times $t_k$.

\begin{figure}[t]
    \centering
    \includegraphics[width=1.00\linewidth]{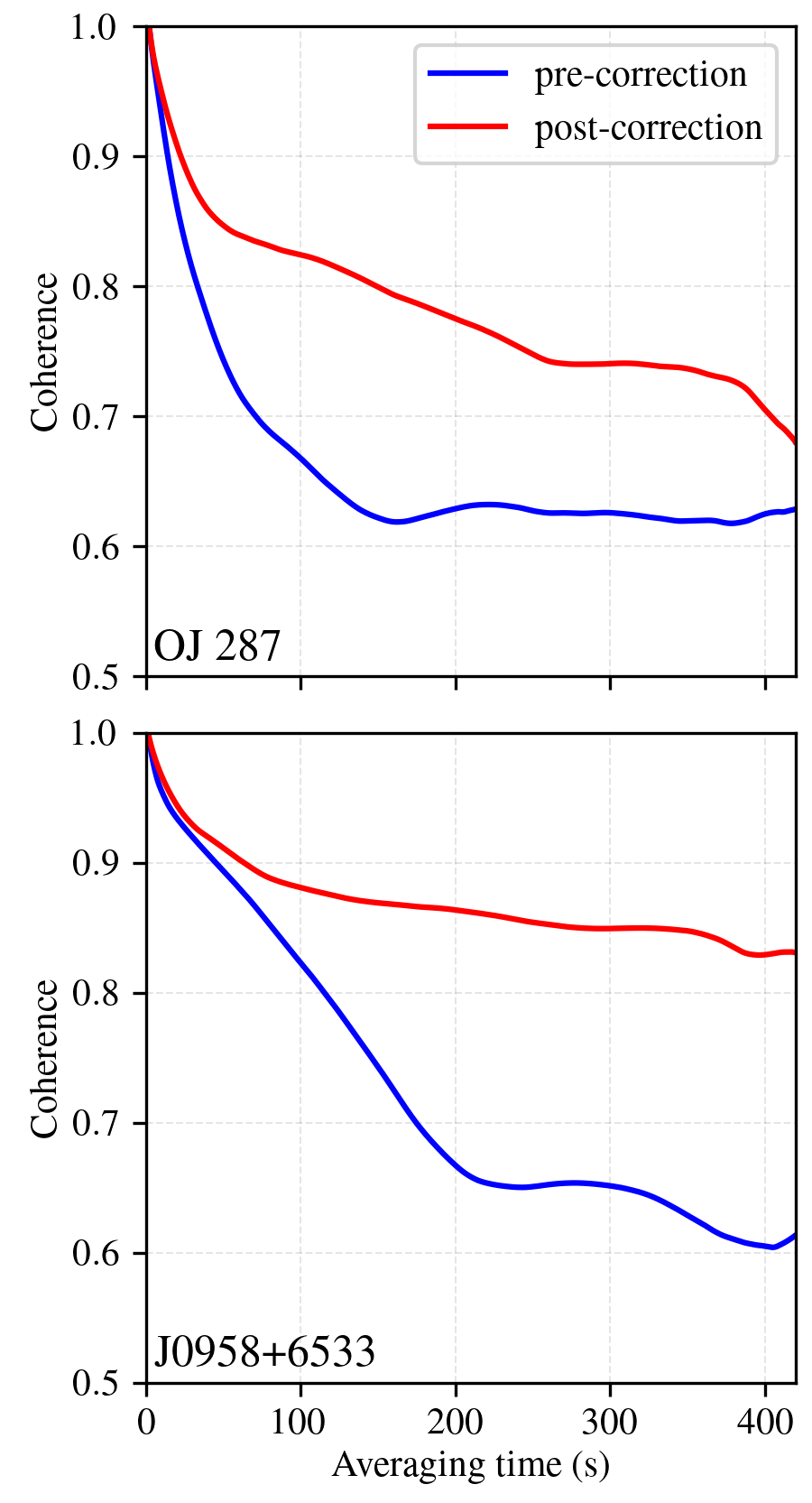}
    \caption{Phase coherence (see \autoref{eqn:Coherence}) as a function of averaging time for the same two scans as shown in \autoref{fig:best-scan}.  The blue curves indicate the 215\,GHz coherence prior to transferring phase corrections from 86\,GHz, and the red curves show the post-correction coherence.}
    \label{fig:coherence}
\end{figure}

We show the coherence of the 215\,GHz visibility phases as a function of averaging time in \autoref{fig:coherence}, both before (in blue; i.e., single-frequency analysis) and after (in red; i.e., dual-frequency analysis) carrying out FPT.  The period of time over which the coherence prior to correction is $>$90\% is $\sim$15\,s for the \oj287 scan and $\sim$40\,s for the \j0958 scan, a difference that is consistent with the factor of $\sim$2 expected from elevation ($\sim$40\,degrees at IRAM for \j0958 compared to $\sim$20\,degrees for \oj287; Hawai`i observed both sources at a similar elevation of $\sim$22\,degrees).  This range of 90\% coherence times is consistent with the values seen in EHT observations at similar frequencies \citep{EHT_M87_Paper2}.  In both scans and at all averaging times, the application of FPT improves the coherence.  For the \j0958 scan, the corrected phase shows a coherence that remains $>$80\% even after averaging over the entire 7-minute scan.

\begin{figure*}[t]
    \centering
    \includegraphics[width=1.00\linewidth]{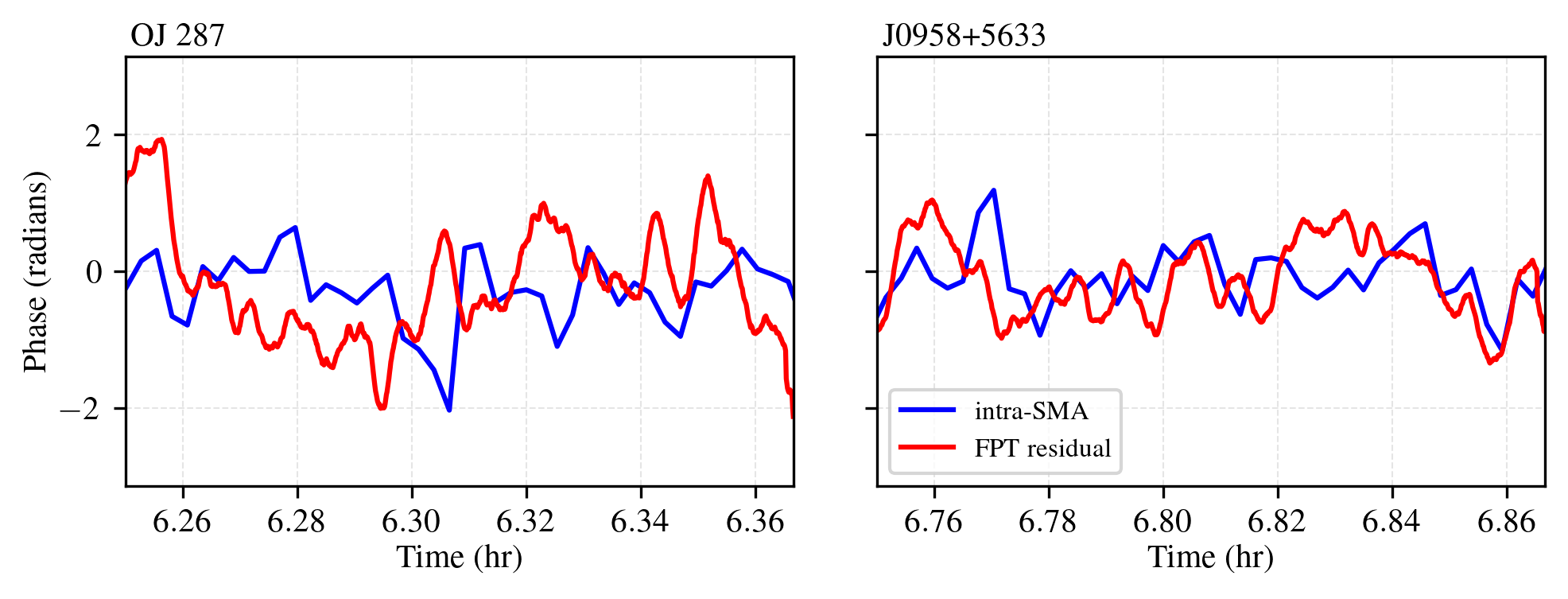}
    \caption{Comparison between the visibility phases on an intra-SMA baseline (in blue) and the phase residual on the VLBI baseline (in red) after carrying out FPT.  The VLBI phase residuals have been coherently averaged over 10\,s to match the SMA cadence.  The intra-SMA phases shown here correspond to the baseline between antennas 2 and 6, as indicated in \autoref{fig:globe}.  The separation between the intra-SMA antennas is $\sim$180\,m, while the separation between the SMA reference antenna and the JCMT is $\sim$160\,m.}
    \label{fig:SMA_phases}
\end{figure*}

\section{Discussion}\label{sec:discussion}

The basic premise underlying FPT is, assuming the non-dispersive nature of the atmosphere, that the atmospheric phase variations observed at one frequency should be proportional to the atmospheric phase variations observed at another frequency, with a proportionality constant equal to the ratio of frequencies.  The correlation observed in \autoref{fig:correlation} indicates that this premise continues to hold for observing frequencies up to 215\,GHz, demonstrating that FPT can be usefully employed at millimeter wavelengths.

However, if the 86 and 215\,GHz phases in our data were perfectly correlated, then the red curves shown in \autoref{fig:coherence} would be horizontal lines with a value of $C=1$; this is clearly not the case.  Furthermore, given our definition for the coherence (see \autoref{eqn:Coherence}), the observed deviation from $C=1$ on finite averaging timescales cannot be attributed entirely to thermal noise in the data.  This fact can also be seen in \autoref{fig:best-scan}: deviations exist between the 215\,GHz and scaled 86\,GHz phase time series that are beyond what thermal noise alone can explain.  Instead, some other source of phase variability is present in the data and acting to drive the coherence below unity.

The most obvious culprit for the excess phase variability is the fact that SMA and JCMT do not share an optical pathway, but are instead separated by a distance of $\sim$160\,m and are thus looking through different patches of the atmosphere.  The phase variations seen by one of these telescopes will not in general be identical to the phase variations seen by the other, and the extent to which their phase variations differ will result in decoherence when averaged;
this effect is further amplified for low elevation observations.
Although we do not have an a priori expectation of how large the differences in observed phase between these two sightlines should be, we can determine such an expectation empirically by comparing the phase difference between SMA and JCMT to the phase difference between two intra-SMA dishes.  \autoref{fig:SMA_phases} shows such a comparison; the red curves show the phase residuals between IRAM$-$SMA and IRAM$-$JCMT (after scaling the latter by the frequency ratio), and the blue curves show the visibility phases measured on an intra-SMA baseline with a separation comparable to the SMA$-$JCMT separation (albeit with a different orientation).

We can see in \autoref{fig:SMA_phases} that the two sets of phases show similar magnitudes and timescales of variation.  For \oj287, the FPT residual phase has a standard deviation of $\sim$0.83 radians, while the intra-SMA phase has a standard deviation of $\sim$0.53 radians. For \j0958, the FPT residual phase and intra-SMA phase have standard deviations of $\sim$0.55 radians and $\sim$0.45 radians, respectively.  A large fraction of the imperfect phase transfer between IRAM$-$SMA and IRAM$-$JCMT can thus plausibly be attributed to the physical separation between SMA and JCMT, amplified by the low-elevation observations. Excess phase variations may also be the result of misalignment between the wind direction and the telescope separation vector \citep[e.g.,][]{Matsushita_2010}.

Although this lack of a shared optical pathway almost certainly explains at least some of the observed excess phase variations, we note that with only a single geometric baseline present in our dataset, we cannot uniquely isolate contributions to the observed phase variability that originate from individual stations.  We thus cannot rule out other sources of phase noise, such as the possibility that the observed variations arise -- in whole or in part -- from some time-variable instrumental instability in one or more stations.  Furthermore, the intra-SMA variations seen in \autoref{fig:SMA_phases} are systematically smaller than the residual VLBI phase variations after FPT, suggesting that there is likely to be an additional source of phase variability in these data beyond what can be explained by just the physical separation of SMA and JCMT.

\section{Summary and conclusions} \label{sec:Summary}

We present results from simultaneous dual-frequency VLBI observations that were carried out at 86\,GHz on the IRAM$-$JCMT baseline and at 215\,GHz on the IRAM$-$SMA baseline on 2024 January 24. The IRAM 30-m observed at both frequencies simultaneously, while the SMA and JCMT acted as a paired antenna on Hawai`i. We observed two bright northern AGN, \j0958 and \oj287, obtaining strong fringes on three scans. The visibility phases at both frequencies exhibit a high degree of correlation that follows the trend expected for non-dispersive atmospheric fluctuations, enabling the first demonstration of FPT up to 1.3\,mm wavelength. Transferring the scaled phases from 86\,GHz to 215\,GHz systematically increases the 215\,GHz coherence on all averaging timescales for both targets.  We attribute residual variations in the post-FPT phases primarily to the $\sim$160\,m physical separation of the JCMT and SMA; these phase variations are comparable in magnitude and timescale to intra-SMA phases between antennas separated by a similar distance.  We expect that observations between two telescopes for which both frequencies share the same optical path would have substantially reduced post-FPT phase residuals. Additional VLBI FPT tests are being planned between stations with true dual-frequency capabilities as they become available.

While there remain challenges in understanding higher-order effects in the atmosphere above spatially separated receivers, the use of the JCMT and SMA as a single dual-frequency antenna in this experiment demonstrates for the first time that FPT is feasible in paired-antenna mode for VLBI \citep{Rioja_2020}. Paired-antenna FPT has practical applications for the participation of connected-element interferometers in multi-frequency VLBI observations.  For instance, while ALMA does not have the capability to conduct multi-band observations \citep[nor will planned upgrades provide this capability; see][]{Carpenter_2023}, dividing the array into sub-arrays that observe at different frequencies could use a paired-antenna FPT calibration scheme.  Such a mode could also form a natural component of the ngVLA's Long Baseline Array (LBA) operation; each of the ten locations of the LBA will host three antennas \citep{Selina_2018}, which could make use of FPT to boost sensitivity at the highest observing frequencies via a paired-antenna approach.  However, we note that because antennas will typically have independent electronics with the potential to result in differential delays and rates, paired-antenna observations are expected to place generically more stringent requirements on instrumental stability than simultaneous multi-band capabilities that share an optical pathway.

As additional (sub)millimeter observatories enable simultaneous observing at the 86 and 230\,GHz bands and beyond, future FPT experiments will greatly benefit from added baselines, where instrumental issues associated with a particular telescope can be more readily identified, quantified, and mitigated.  Future observations tuning the observing bands to an exact integer frequency ratio could quantify the improvement in the quality of the FPT when phase ambiguities are entirely avoided. FPT observations, particularly of fainter targets, would also benefit from a priori phase calibration at the instrument, such as phase-cal tone injection to flatten the bandpass. Longer scan lengths, potentially to tens of minutes, would further test the coherence improvement from FPT beyond typical timescales for (sub)millimeter observing. Finally, the interleaving of two nearby targets would allow for a test of another variant of FPT called ``source frequency phase referencing'' (SFPR): this technique allows for the removal of remaining FPT dispersive residual terms  -- including instrumental terms --  using the secondary target and astrometrically registers the structures of the primary target across the observed frequencies \citep{Dodson_2009,Rioja_2011,Rioja_2015,Yoon_2018,Wu_2018,Wu_2021,Wu_2023}.

The observations presented here retire a key uncertainty associated with the FPT technique, demonstrating that the technique remains viable up to an observing wavelength of $\sim$1\,mm, even with paired antennas, and paving the way for substantial sensitivity boosts for high-frequency VLBI.

\software{\texttt{AIPS} \citep{aips,greisen2011fits}, \texttt{AllanTools}\footnote{\url{https://github.com/aewallin/allantools}}, \texttt{HOPS} \citep{Whitney_2004,lindyhops}, \texttt{matplotlib} \citep{Hunter_2007}, \texttt{numpy} \citep{numpy_2011,numpy_2020}, \texttt{pandas} \citep{pandas}}

\facility{SMA, JCMT, IRAM 30-m}

\acknowledgments

We are indebted to the efforts of JCMT/EAO staff members Izumi Mizuno, Per Friberg, Mark G. Rawlings, and Ryan Foley, without whom the observations presented in this paper would not have been possible.

The photograph of Maunakea used in \autoref{fig:globe} is taken from Google Earth.

Support for this work was provided by the NSF through grants AST-1935980 and AST-2034306, and by the Gordon and Betty Moore Foundation through grants GBMF5278 and GBMF10423. This work has been supported in part by the Black Hole Initiative at Harvard University, which is funded by the John Templeton Foundation (grants 60477, 61479, and 62286) and the Gordon and Betty Moore Foundation (grant GBMF8273).  This work is based on observations carried out with the IRAM 30m telescope. IRAM is supported by INSU/CNRS (France), MPG (Germany) and IGN (Spain). S.\,I. was supported by Hubble Fellowship grant HST-HF2-51482.001-A awarded by the Space Telescope Science Institute, which is operated by the Association of Universities for Research in Astronomy, Inc., for NASA, under contract NAS5-26555. I. M-V. acknowledges funding support from Projects PID2022-140888NB-C22 and PID2019- 108995GB-C22 (Ministerio de Ciencia, Innovacion y Universidades), and Project ASFAE/2022/018 (Generalitat Valenciana). G.-Y.\,Z. receives support from the M2FINDERS project that is funded by the European Research Council (ERC) under the European Union's Horizon 2020 research and innovation program (grant agreement No. 101018682). R.-S.\,L. is supported by the National Science Fund for Distinguished Young Scholars of China (Grant No. 12325302) and the Shanghai Pilot Program for Basic Research, CAS, Shanghai Branch (JCYJ-SHFY-2021-013). The Submillimeter Array is a joint project between the Smithsonian Astrophysical Observatory and the Academia Sinica Institute of Astronomy and Astrophysics and is funded by the Smithsonian Institution and the Academia Sinica. We recognize that Maunakea is a culturally important site for the indigenous Hawai`ian people; we are privileged to study the cosmos from its summit. The James Clerk Maxwell Telescope is operated by the East Asian Observatory on behalf of The National Astronomical Observatory of Japan; Academia Sinica Institute of Astronomy and Astrophysics; the Korea Astronomy and Space Science Institute; the National Astronomical Research Institute of Thailand; Center for Astronomical Mega-Science (as well as the National Key R\&D Program of China with No. 2017YFA0402700). Additional funding support is provided by the Science and Technology Facilities Council of the United Kingdom and participating universities and organizations in the United Kingdom and Canada. Nāmakanui, which is the receiver used for 86\,GHz observing at JCMT, was constructed and funded by ASIAA in Taiwan, with funding for the mixers provided by ASIAA and at 230\,GHz by EAO. The Nāmakanui instrument is a backup receiver for the GLT. The authors wish to recognize and acknowledge the very significant cultural role and reverence that the summit of Maunakea has always had within the indigenous Hawai`ian community.  We are most fortunate to have the opportunity to conduct observations from this mountain.
We acknowledge the efforts of Dan Bintley, Jamie Cookson, Paul T. P. Ho, Shaoliang Li, Kuan-Yu Liu, Akil Marshall, and Andrea McCl\"oskey in making the JCMT observations reported in this paper possible.

\bibliography{references, EHTCPapers}{}
\bibliographystyle{aasjournal}

\end{document}